\documentstyle[preprint,aps]{revtex}
\begin{document}
\bibliographystyle{plain}
\title{
Susceptibility of the one-dimensional, dimerized Hubbard model
}
\vskip0.5truecm
\author {Fr\'ed\'eric Mila$^{a}$, Karlo Penc$^{b(*)}$}
\vskip0.5truecm
\address{
     $(a)$ Laboratoire de Physique Quantique, Universit\'e Paul Sabatier\\
     31062 Toulouse (France)\\
$(b)$ Tokyo Institute of Technology, Department of Physics\\
Oh-okayama, Meguro-ku, Tokyo 152 (Japan) }
\maketitle

\begin{abstract}
We show that the zero temperature susceptibility of
the one-dimensional, dimerized Hubbard model at quarter-filling
can be accurately determined on the basis of exact diagonalization
of small clusters. The best procedure is to perform a finite-size scaling
of the spin velocity $u_\sigma$, and to calculate the susceptibility from
the Luttinger liquid relation $\chi=2/\pi u_\sigma$.
We show that these results are reliable by comparing them with the
analytical results that can be obtained in the weak and strong coupling limits.
We have also used quantum Monte Carlo simulations to calculate
the temperature dependence of the susceptibility for parameters that should be
relevant to the Bechgaard salts. This shows that, used together, these
numerical techniques are able to give precise estimates of the low temperature
susceptibility of realistic one-dimensional models of correlated electrons.
\end{abstract}
\vskip .1truein

\vskip 0.1 truein
(*) On leave from Research Institute for Solid State Physics, Budapest,
Hungary.
\newpage
The properties of correlated electrons are fairly well understood in one
dimension\cite{solyom,emery}.
The low-energy physics of systems with low lying charge and spin excitations is
well described by the Tomonaga-Luttinger fixed-point, while the properties of
systems with a gap in the charge or in the spin sector can be inferred from
those of the half-filled Hubbard model or of the Luther-Emery model
respectively. When it comes to a precise understanding of the one-dimensional
properties of real materials, like the organic conductors, this is not
sufficient however. The ultimate goal is to understand both the high and low
energy properties of a given material in terms of a few parameters describing
the basic electronic processes, namely hopping integrals and Coulomb
repulsions.
For instance, for the Bechgaard salts, a good candidate to describe the
electronic properties is given by the quarter-filled, dimerized Hubbard
model\cite{mila1}
defined by the Hamiltonian
\begin{eqnarray}
H = -t_1 \sum_{i {\rm even}, \sigma} (c_{i,\sigma}^\dagger
c_{i+1,\sigma}^{\vphantom{\dagger}} + {\rm h.c.})
      -t_2 \sum_{i {\rm odd}, \sigma} (c_{i,\sigma}^\dagger
      c_{i+1,\sigma}^{\vphantom{\dagger}} + {\rm h.c.})
      + U \sum_{i} n_{i\uparrow}n_{i\downarrow}
\end{eqnarray}
The parameters of this model are: i) a hopping integral $t_1$ for the short
bonds; ii) a hopping integral $t_2$ ($\le t_1$) for the long bonds; iii) an
on-site repulsion $U$. In the following, energies will be measured in units of
$t_1$, and the basic dimensionless parameters are $t_2/t_1$ for the
dimerization
and $U/t_1$ for the Coulomb interaction.
To check whether this is a fair description of the low energy properties,
one needs reliable estimates of the quantities that
are accessible experimentally. For instance, the
activated behaviour of the resistivity at low temperature in (TMTTF)$_2$PF$_6$
is believed to be due to a small gap in the charge spectrum\cite{wzietek}.
To get a
quantitative estimate of the charge gap in the model of Eq. (1) is a
difficult -- but not hopeless -- problem that we have addressed in a previous
paper\cite{penc}. Another useful
probe of the low energy physics is provided by the static susceptibility,
which has been measured quite extensively as a function of temperature
for the Bechgaard
salts\cite{wzietek}. The
interpretation that has been proposed so far is based on analytical results
obtained in the weak-coupling limit of the non-dimerized model.
This is probably a good approximation for (TMTSF)$_2$PF$_6$ because the
dimerization is small, but not for (TMTTF)$_2$PF$_6$, in which case the ratio
$t_2/t_1$ is about 0.7, and a calculation that takes the dimerization
into account
should be done before a reliable interpretation of the data can be tried.
Dimerization is actually a very common phenomenon in one-dimensional
conductors,
and such a calculation is likely to be useful in other contexts as well.

Surprisingly enough, it seems that there has not been
any serious attempt
at calculating the
susceptibility of a dimerized model. Even for the Hubbard model, to which
most of the effort has been devoted because it is soluble with the Bethe
ansatz\cite{lieb}, the information that can be found in the
litterature is not complete.
The susceptibility has been calculated at zero temperature for any density
by Shiba\cite{shiba} with the Bethe ansatz equations, but
its temperature dependence
could be determined along these lines only at half-filling
\cite{pincus}.
As a consequence,
Torrance et
al\cite{torrance} had to use approximate results obtained in the limiting cases
of weak and strong on-site repulsion
to analyze the susceptibility of TTF-TCNQ
because the band-filling is not 1 but 0.59.
A few years later,
the temperature dependence of the susceptibility for the quarter-filled case
and
for $U/t=4$ was determined through Monte Carlo simulations by Hirsch and
Scalapino\cite{hirsch1}, but this was just intended as an illustration,
and the precision was not good enough at low temperature for these results to
be
used in the interpretation of experimental results. In fact, the only attempt
at
calculating the temperature dependence of the susceptibility away from
half-filling is due to Bourbonnais\cite{bourbonnais}.
His calculation is based on g-ology. The
fact that the system is not half-filled is taken into account by neglecting the
umklapp term. While this calculation seems reliable for not too large values of
$U/t$ and at low temperatures, it clearly fails at higher temperature, the
susceptibility diverging instead of vanishing as $1/T$.

In the present paper, we show that it is now possible to get accurate results
for both the zero-temperature value and the temperature
dependence of the susceptibility for one-dimensional models that do not have an
exact solution by
using standard numerical techniques together with analytical results in various
limits. We have decided to concentrate on the model of Eq. (1) because
it has a direct relevance to the Bechgaard salts, for which extensive
experiemental data are available, but the same approach can in principle be
used
for any model, with hopefully the same success.

Numerically, there are a priori several ways to calculate the susceptibility.
One
possibility is to go back to the original definition of the susceptibility as
the derivative of the energy with respect to the magnetization evaluated for
vanishing magnetization. This is not so
accurate however because, with finite clusters,
one has only values of the energy for discrete values
of the total magnetization, and one first has to do a linear fit of the energy
as a function of magnetization for small values of the
magnetization. This procedure turns out to be rather arbitrary for the
model of Eq. (1) when the repulsion is large. For one-dimensional systems,
there
is another way to calculate the susceptibility. Unless there is gap in the spin
sector, the low energy spin excitations of electrons
interacting through repulsive interactions
can be described by a Luttinger liquid with a velocity
$u_\sigma$, and the susceptibility can be obtained as $\chi = 2/\pi
u_\sigma$\cite{schulz}.
One route to the susceptibility is then to determine $u_\sigma$ numerically
from
exact diagonalization fo small clusters. There are again several possibilities
to extract this velocity. When $t_2/t_1\neq 1$, we have shown\cite{penc} that
there is a
gap in the charge sector of the model of Eq. (1). The ground-state energy is
then expected to scale with
the number of sites $L$ as\cite{frahm}
\begin{eqnarray}
{E_0 \over L} = \epsilon - {\pi u_\sigma c \over 6 L^2} + o\left({1 \over
L^2}\right)
\end{eqnarray}
where $\epsilon$ is the energy density in the thermodynamic limit, and where
the central charge $c$ should be equal to 1. In the present case, this
cannot be a good way to determine $u_\sigma$. First,
there are logartihmic corrections to the coefficient of the $1/L^2$ term which
makes the estimate based on small clusters not very accurate.
Second, and more importantly, this scaling behaviour is probably not satisfied
fot the sizes we can study with exact diagonalization. The reason is quite
simple: When the dimerization is
very small, the gap $\Delta_c$ in the charge sector is also very small.
Now, the scaling
of Eq. (2)
can be observed only when the size of the systems is larger than the
correlation
length associated with this gap, which is given by $\xi \sim u_c / \Delta_c$,
where $u_c$ is the slope of the dispersion of the charge excitations for
energies
larger than the gap. If the systems are small, the scaling of the ground-state
will probably be closer to the formula for systems without a gap in the charge
sector, which is similar to Eq. (2), but with $u_\sigma$ replaced by $u_\sigma
+
u_c$, and Eq. (2) would provide a totally unreliable estimate of
$u_\sigma$.

So it seems that the best way to determine $u_\sigma$ is to look at the
spectrum directly, and to determine the slope of the spin excitations. More
precisely, the finite-size estimate of $u_\sigma$ is given by
\begin{eqnarray}
u_\sigma(L) = {E(k_0+2\pi/L;S=1;L)-E(k_0;S=0;L) \over 2\pi/L}
\end{eqnarray}
where $k_0$ is the momentum of the ground-state, and where $E(k;S;L)$ is the
lowest energy in the subspace of states of momentum k and total spin S
for a system of
size L. This
quantity is expected to go to the value $u_\sigma$ of the infinite system with
dominant corrections of order $1/L$. We have calculated $u_\sigma(L)$
for systems with 8, 12 and 16 sites, and for several values of $t_2/t_1$ and
$U/t_1$. This scaling was in most cases already very accurately satisfied.
Small but not negligible deviations occured for large values of $U/t_1$. As the
form of the next-to-leading order corrections is not clear,
we have used the systems with 12
and 16 sites to perform the $1/L$ extrapolation. The results for $\chi$
are given in Fig. 1.

We have performed several checks to convince ourselves that these
estimates of the zero-temperature susceptibility were accurate. First,
when $t_2/t_1=1$, the model of Eq. (1) is nothing but the Hubbard model.
Solving numerically the equations of the Bethe ansatz,
Shiba\cite{shiba} calculated the susceptibility
for several values of the density and of $U$. Our results are in perfect
agreement with the results he quoted for quarter-filling.

In the weak-coupling limit, one can use $g$-ology to determine the
corrections to
the spin velocity. This leads to the following expression for the
susceptibility:
\begin {eqnarray}
\chi = {2 \over \pi v_F} \left(1+ {U \over 2\pi v_F} \right)
\end{eqnarray}
and $v_F=2t_1t_2/\sqrt{t_1^2+t_2^2}$ for the model of Eq. (1). This is again
in very
good agreement with the slopes we have obtained numerically for small values of
$U/t_1$.

Given the uncertainty of our numerical procedure for large values of $U/t_1$,
it
is also very important to check the results that we have obtained in that
limit.
 This
turns out to be possible along the lines we used to calculate the $1/U$
corrections to the charge gap\cite{penc}. When $U/t_1$ is very large, the
ground-state
is approximately given by the product of the spinless fermion Fermi sea
with the ground-state wave function of the spin-1/2 Heisenberg model,
the number of spins being
equal to the number of electrons in the original model\cite{ogata}. The
coupling between the
spins is described by an effective exchange integral $J_{\rm eff}$ given by
\begin{equation}
 J_{\rm eff}=\frac{t_1^2 + t_2^2} {2U}-
         \frac{2 (t_1^2 - t_2^2)}{\pi^2 U}   K^2\left({2
         \sqrt{t_1 t_2} \over t_1+t_2 }\right)
\end{equation}
where
\begin{equation}
K(q) = \int^{\pi/2}_0 \frac{d\phi} {\sqrt{1-q^2\sin \phi ^2}}
\end{equation}
is an elliptic integral of the first kind.
The susceptibility is given in terms of $J_{\rm eff}$ by\cite{ogata}
\begin{equation}
\chi = \frac {1} {\pi^2 J_{\rm eff}}
\end{equation}
This result can be seen as a generalization of the result obtained by Klein and
Seitz\cite{klein} for the Hubbard model.
To compare with our results, we have plotted $\chi/U$ as a function of $1/U$,
the result for $1/U=0$ being deduced from the previous equations. The agreement
is satisfactory. For large $U$, our numerical results for $\chi/U$ are slightly
scattered, but their are consistent with the infinite $U$ result. We consider
this agreement as a strong support in favour of this procedure to estimate the
zero temperature susceptibility of one-dimensional models of correlated
electrons for large values of the repulsion.

Quite independently from these results,
information on the susceptibility at moderate to high temperatures can
be obtained from Monte Carlo
simulations. Experimentally, the temperature
dependence of the susceptibility is usually known quite accurately, and its
interpretation is not ambiguous in the sense that it gives access to the ratio
$U/t$, which is a mesure of the size of correlations, even if one does not know
the value of the hopping integrals accurately. This should be contrasted to the
low temperature value of the susceptibility: To give information on the size of
correlation, this quantity has to be compared with estimates for non
interacting
electrons, and, for molecular conductors, these estimates are not reliable
if they are deduced solely from band structure calculations.

The world-line algorithm of Hirsch et al\cite{hirsch2} is known to be
very convenient
for one-dimensional models, and we have used it to determine the
temperature dependence of the susceptibility for two sets of parameters:
$U/t_1=4$, $t_2/t_1=1$, which should be reasonable for (TMTSF)$_2$PF$_6$, and
$U/t_1=8$, $t_2/t_1=.7$, which should be reasonable for (TMTTF)$_2$PF$_6$. In
their original study of the extended Hubbard model, Hirsch et al worked at
temperatures such that the finite size effects were smaller than the
statistical
errors. In our case, we need to go to temperatures that are small enough to
make the link with the zero temperature results obtained with exact
diagonalization. So we had to obtain results with statistical errors
small enough
to allow meaningful extrapolations.
More precisely, we had to do three extrapolations. First, in the algorithm we
have used, the magnetization is a conserved quantity. So, to get the static
susceptibility, one has to extrapolate the finite $q$ results to get the
zero-$q$ value of the susceptibility. Practically, this is done by fitting the
results at small $q$ with a parabola. Then, there is the systematic error
due to the Trotter decomposition which goes as $\Delta \tau ^2$, and  which we
eliminated by fitting the results obtained for different numbers of temperature
slices. Finally, for the lowest temperatures studied,
the finite-size effects were not negligible, and, by analogy
with the procedure used for the zero-temperature susceptibility, we performed a
linear fit in $1/L$ of the results obtained for different sizes to go to the
thermodynamic limit. The error bar
depicted on the figure is the largest of the statistical errors
obtained for the various sizes and time-slice numbers for a given value of the
temperature.
The results are shown in Fig. 3. The agreement with the zero-temperature result
is very good. This is particularly satisfactory for the case $U/t_1=8$,
$t_2/t_1=.7$, because this was already in the region where the finite-size
corrections to $u_\sigma$ were not purely $1/L$ for the small systems we could
study with exact diagonalization.

As far as the Bechgaard salts are concerned, we will limit ourselves to a few
remarks. A more complete account of these results, together with the
interpretation of several other experimental data, will be given
elsewhere. Let
us note for the moment that the present results roughly confirm the
interpretation given by Wzietek et al\cite{wzietek} for (TMTSF)$_2$PF$_6$:
Our result for the
temperature dependence of the susceptibility for $U/t_1=4$, $t_2/t_1=1$ is
consistent with the results of Bourbonnais. For (TMTTF)$_2$PF$_6$, our results
suggest that the dimerization reduces the temperature dependence of the
susceptibility, so that the analysis of Wzietek et al, which did not take the
dimerization into account, leads to an underestimate of $U/v_F$ in that
compound.

In conclusion, we have shown that it is possible to obtain accurate results
concerning the susceptibility of one-dimensional models of correlated electrons
that cannot be solved exactly by Bethe ansatz by using numerical techniques and
approximate methods in various limits. A finite-size scaling of the spin
velocity deduced from the spectrum obtained by exact diagonalization provides
good values of the zero-temperature susceptibility, while standard Monte Carlo
simulations give information on the temperature dependence that seems accurate
enough to allow an interpretation of the experimental data obtained on the
Bechgaard salts. It is our hope that this work will encourage experimentalists
to try to interprete the low energy results they can obtain for one-dimensional
conductors in terms of microscopic models.

We acknowledge useful discussions with L. Brossard, D. J\'erome, D. Poilblanc,
J.-P. Pouget, S. Sorella, P. Wzietek and T. Ziman. We are
especially indebted to M. Dzierzawa for providing us with a Quantum Monte Carlo
program that we could adapt to calculate the finite temperature susceptibility.
The numerical simulations were performed on the C98 of the IDRIS (France).

\begin{figure}
\caption{ Zero temperature susceptibility
as a function of $U/t_1$ for several
values of the dimerization: a) $t_2/t_1$ = 0.4, 0.5, ..., 1 from top to bottom;
b) $t_2/t_1$ = 0.1, 0.2, 0.3 from top to bottom. }
\end{figure}

\begin{figure}
\caption{ $\chi t_1/U$ as a function of $t_1/U$ for several
values of the dimerization: a) $t_2/t_1$ = 0.4, 0.5, ..., 1 from top to bottom;
b) $t_2/t_1$ = 0.1, 0.2, 0.3 from top to bottom.  }
\end{figure}

\begin{figure}
\caption{ Temperature dependence of the susceptibility for: a) $U/t_1=4$,
$t_2/t_1=1$; b)
$U/t_1=8$, $t_2/t_1=0.7$.}
\end{figure}

\end{document}